\documentclass[aps,prl,preprintnumbers,amsmath,amssymb,twocolumn,superscriptaddress,showpacs,floatfix]{revtex4}

\usepackage{graphicx}
\usepackage{dcolumn}
\usepackage{bm}

\begin{document}
\title{Globally Polarized Quark-gluon Plasma in Non-central $A+A$ Collisions}

\author{Zuo-Tang Liang}
\affiliation{Department of Physics, Shandong University, Jinan, 
Shandong 250100, China}

\author{Xin-Nian Wang}
\affiliation{Nuclear Science Division, MS 70R0319,
Lawrence Berkeley National Laboratory, Berkeley, California 94720}
\affiliation{Department of Physics, Shandong University, Jinan, 
Shandong 250100, China}

\date{October 18, 2004}

\preprint{LBNL-56383}

\begin{abstract}
Produced partons have large local relative orbital 
angular momentum along the direction opposite to the 
reaction plane in the early stage of non-central heavy-ion collisions.
Parton scattering is shown to polarize quarks along 
the same direction due to spin-orbital coupling. Such global quark 
polarization will lead to many observable consequences, such as
left-right asymmetry of hadron spectra, global transverse polarization of 
thermal photons, dileptons and hadrons. Hadrons from the decay
of polarized resonances will have azimuthal asymmetry similar to
the elliptic flow. Global hyperon polarization is studied within 
different hadronization scenarios and can be easily tested.
\end{abstract}

\pacs{25.75.-q, 13.88.+e, 12.38.Mh, 25.75.Nq}

\maketitle

Strong transverse polarization of hyperons has been observed
in unpolarized $p+p$ and $p+A$ collisions since 
the 1970's\cite{PhData}.
Given the beam and hyperon momenta $\vec{p}$ and $\vec{p}_H$, 
hyperons produced in the beam fragmentation region 
are found transversely polarized in the direction perpendicular 
to the hyperon production plane,
$\vec{n}_H= \vec{p} \times \vec{p}_H/|\vec{p}\times\vec{p}_H|$.
Polarizations of $\Lambda$, $\Xi$ and $\bar\Xi$ are negative while
$\Sigma$ and $\bar\Sigma$'s are positive. In the meantime, $\bar\Lambda$
and $\Omega$ are not transversely polarized.
Although the origin for such striking transverse hyperon polarization 
is still in debate, one can relate it to the single-spin left-right 
asymmetries observed in hadron-hadron collisions with transversely 
polarized beam \cite{LB97}, which in turn can be attributed to the 
orbital angular momenta (o.a.m.) of the valence quarks in a 
polarized nucleon \cite{LM92,LB00,Troshin:1996hd}, or
fragmentation functions of transverse polarized 
quarks \cite{Collins:1992kk} as well as
twist-3 parton correlations in nucleons \cite{Qiu:1998ia}. 
It has also been suggested \cite{stock,Panagiotou:1986zq} that hyperon 
polarization could disappear due to the formation of QGP.


In this Letter, we show that parton interaction in {\it non-central} 
heavy-ion collisions leads to a global quark polarization 
along the opposite direction of the reaction plane, 
\begin{equation}
\vec{n}_b = \vec{p} \times \vec{b}/|\vec{p}\times\vec{b}|,
\label{eq:plane}
\end{equation}
as determined by the nuclear impact parameter $\vec{b}$. This global
polarization is essentially a local manifestation 
of the global angular momentum of the colliding system through
interaction of spin-orbital coupling in QCD. It
will have far reaching consequences in non-central heavy-ion 
collisions, such as left-right asymmetry of hadron spectra in the
reaction plane, global transverse polarization of direct photons, 
dileptons and hadrons with spin. Within different hadronization
scenarios,
we will discuss hyperon polarization as a result of the global quark 
polarization. Possible contribution from final state hadronic 
interaction will also be discussed.

\begin{figure}[htbp]
\resizebox{3.0in}{2.4in}{\includegraphics{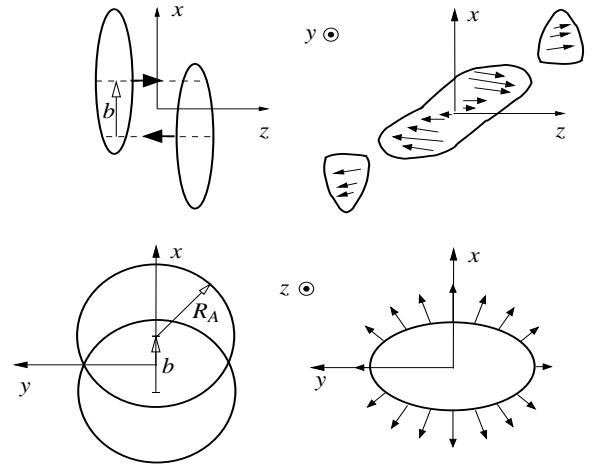}}
\caption{Illustration of non-central heavy-ion collisions with
impact parameter $\vec{b}$. The
global angular momentum of the produced matter is along $-\hat{y}$,
opposite to the reaction plane.}
\label{fig1}
\end{figure}

Let us consider two colliding nuclei with the beam projectile
moving in the direction of the $z$ axis, as illustrated in Fig.~\ref{fig1}.
We define the impact parameter $\vec{b}$ (along $\hat{x}$) as 
the transverse distance of the projectile from the target nucleus 
and the reaction plane as given by $\vec{n}_b$ (along $\hat{y}$) 
in Eq.~(\ref{eq:plane}). Partons produced in the overlapped 
region of the collision will carry a global angular momentum 
along the direction opposite to the reaction plane ($-\hat{y}$). 
A thermalized QGP requires final state parton interaction.
Given the nature of partonic interaction at high energy, the 
global angular momentum would never lead to a collective rotation
of the system. It will, however, be
manifested in the finite transverse (along $\hat{x}$) gradient 
of the average longitudinal momentum $p_z(x,y,b)$ per produced parton.
We assume for the moment that $p_z(x,y,b)$ is independent
of the longitudinal position and is just an average value.
Given the range of interaction $\Delta x$, two colliding
partons will have relative longitudinal momentum 
$\Delta p_z=\Delta x dp_z/dx$ with o.a.m.
$L_y \sim -\Delta x\Delta p_z$ along
the direction of $\vec{n}_b$. This relative o.a.m. $L_y$ 
will lead to global quark polarization due to spin-orbital coupling.

The initial collective longitudinal momentum can be calculated 
as the total momentum difference between participant projectile
and target nucleons, whose transverse distributions (integrated
over $y$) are,
\begin{eqnarray}
\frac{dN_{\rm part}^{P,T}}{dx}&=&\frac{3A}{2\pi R_A}\left\{
\tilde {y}_{\rm max}\sqrt{1-\tilde{y}_{\rm max}^2
-(\tilde{x} \mp \tilde{b}/2)^2} \right. \nonumber\\
& &\hspace{-0.55in}+\left. [1-(\tilde{x} \mp \tilde{b}/2)^2]
\arcsin
\left[\tilde {y}_{\rm max}/\sqrt{1-(\tilde{x} \mp \tilde{b}/2)^2}\right]
\right\}; \nonumber \\
\tilde {y}_{\rm max}&=&\min\left\{\sqrt{1-(\tilde{x}+\tilde{b}/2)^2},
\sqrt{1-(\tilde{x} - \tilde{b}/2)^2}\right\},
\end{eqnarray}
where $\tilde{x}=x/R_A$, $\tilde{b}=b/R_A$ and $R_A=1.12A^{1/3}$
is the nuclear radius in a hard-sphere distribution.

\begin{figure}[htbp]
\resizebox{3.1in}{2.1in}{\includegraphics{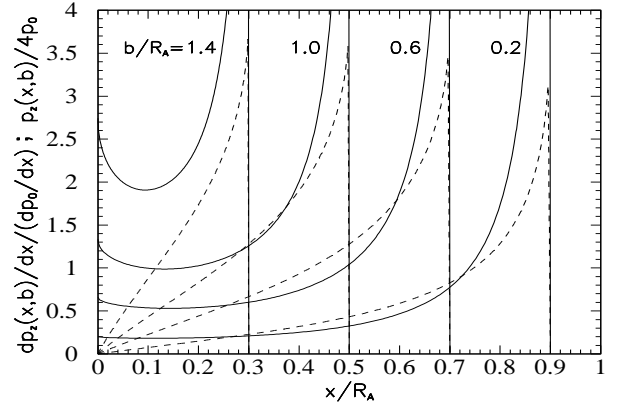}}
\caption{$(dp_z/dx)/(dp_0/dx)$ (solid) and  $p_z(x,b)/4p_0$ 
(dashed) as functions of $x/R_A$ for
different values of $b/R_A$.}
\label{fig2}
\end{figure}

Since the measured total multiplicity in $A+A$ collisions 
is proportional to the number of participant nucleons \cite{phobos2}, 
we can assume the same for the produced partons 
with a proportionality $c(s)$. The average collective 
longitudinal momentum per parton is then,
\begin{equation}
p_z(x,b)=\left(\frac{\sqrt{s}}{2c(s)}\right)\frac{dN_{\rm part}^P/dx 
- dN_{\rm part}^T/dx}
{dN_{\rm part}^P/dx + dN_{\rm part}^T/dx}.
\end{equation}
The distribution $p_z(x,b)$ is an odd 
function in both $x$ and $b$ and therefore vanishes at $x=0$ or $b=0$.
As shown in Fig.~\ref{fig2} (as dashed lines) in 
units of $4 p_0\equiv 4 \sqrt{s}/2c(s)$ 
and as a function of $\tilde{x}$ for different values of $\tilde{b}$, 
it is a monotonically increasing function of $\tilde{x}$ until the edge of 
the overlapped region $|\tilde{x}\pm \tilde{b}/2|=1$ 
where it drops to zero. The transverse gradient $dp_z/dx$,
shown as solid lines in unit of 
$dp_0/dx\equiv \sqrt{s}/2c(s)R_A$, is an even function of $x$ and 
vanishes at $b=0$. It increases almost linearly with $b$ for small 
and intermediate values of $b$. Except the singular behavior at 
the boundary of the overlapped region which is caused by the 
assumed hard-sphere nuclear distribution, $dp_z/dx$
is approximately uniform across the transverse $x$-direction.
The classical relative o.a.m.
is then $L_y \equiv -(\Delta x)^2dp_z/dx$ for
partons separated by $\Delta x$ in the transverse direction.
Due to transverse expansion, $dp_z/dx$ will decrease with time
according to $(R_A-b/2)/[R_A-b/2+v_x(b)(\tau-\tau_0)]$,
where $v_x(b)$ is the transverse flow velocity in 
the reaction plane.

In $Au+Au$ collisions at $\sqrt{s}=200$ GeV, the number of
charged hadrons per participant nucleon is about 15\cite{phobos2}.
Assuming the number of partons per (meson dominated) hadron is about
2, then $c(s)\simeq 45$. Given $R_A=6.5$ fm, $dp_0/dx \simeq 0.34 $ GeV/fm
and $L_0\equiv -(\Delta x)^2 dp_0/dx \simeq 1.7$ for $\Delta x=1$ fm.

We can relax the approximation of uniform distribution 
of $p_z(x,b)$ in the longitudinal direction by identifying
pseudo-rapidity with the spatial rapidity $\eta=0.5\ln(t+z)/(t-z)$.
According to experimental studies of hadron
production in $p+A$ and $A+B$ collisions, 
the collective longitudinal momentum $p_z$ and $dp_z/dx$ are distributed
across a broad range of rapidity and peaks in the forward (backward)
region for a given layer of dense matter at positive (negative) $x$.
The position of the peak in rapidity should increase with $|x|$. Note
that even though $p_z(x,b)$ vanishes at around $x=0$, $dp_z/dx$ and
$L_y$ are still finite for $b\neq 0$ as shown in Fig.~\ref{fig2}.
Averaging over the $x$-direction will result in finite $dp_z/dx$ 
and $L_y$ around central rapidity region in non-central 
heavy-ion collisions.

To study quark polarization due to parton collisions 
with a fixed direction of o.a.m.,
we consider quark scattering with fixed impact parameter $\vec{x}_T$. 
For given initial relative momentum $(E,\vec{p})$
and final spin $\lambda/2$ of the quark along $\vec{n}_b$, 
the cross section is,
\begin{eqnarray}
& & \frac{d\sigma_{\lambda}}{d^2x_T}=
C_T\int \frac{d^2q_T}{(2\pi)^2}
\frac{d^2k_T}{(2\pi)^2} e^{i(\vec{k}_T-\vec{q}_T)\cdot \vec{x}_T}
{\cal I}_{\lambda}(\vec{q}_T,\vec{k}_T,E),\nonumber \\
& &{\cal I}_{\lambda}
=\frac{g^2}{2(2E)^2}\bar{u}_{\lambda}(p_q)A\hspace{-5pt}\slash (\vec{q}_T) 
(p\hspace{-4.5pt}\slash+m_q)
A\hspace{-5pt}\slash (\vec{k}_T) u_{\lambda}(p_k),
\end{eqnarray}
within the screened potential model \cite{gw93},
where $A_0(q_T)=g/(q_T^2+\mu^2)$ is the screened static potential
with Debye screen mass $\mu$, $C_T$ 
is the color factor associated with the target. 
Average over spin polarization is implied for the initial quark 
and $\vec{p}_{q(k)}=\vec{p}+\vec{q}_T(\vec{k}_T)$ is the 
final quark momentum. For small angle 
scattering, $q_T,k_T \sim \mu \ll E$,
\begin{equation}
\frac{{\cal I}_{\lambda}}{g^2}\approx
\frac{1}{2}A_0(q_T)A_0(k_T)\left[1-i \lambda\frac{
(\vec{q}_T-\vec{k}_T)\cdot (\vec{n}_b\times\vec{p})}
{2E(E+m_q)}
\right].
\end{equation}
The first term gives the unpolarized cross section
\begin{equation}
\frac{d\sigma}{d^2x_T}\equiv \frac{d\sigma_+}{d^2x_T}
+\frac{d\sigma_-}{d^2x_T}=4C_T \alpha_s^2 K_0^2(\mu x_T),
\end{equation}
or $d\sigma/dq_T^2=C_T4\pi\alpha_s^2/(q_T^2+\mu^2)^2$ in momentum space.
The second term gives rise to a polarized cross section
$d\Delta\sigma/d^2x_T \equiv d\sigma_+/d^2x_T-d\sigma_-/d^2x_T$,
\begin{equation}
\frac{d\Delta\sigma}{d^2x_T}=
-\mu \frac{\vec{p}\cdot(\hat{x}_T \times \vec{n}_b)}
{E(E+m_q)}
4C_T\alpha_s^2K_0(\mu x_T)K_1(\mu x_T),
\end{equation}
where $\hat{x}_T=\vec{x}_T/x_T$ and
$K_n$'s are modified Bessel functions.
It is evident that parton scattering polarizes
quarks along the direction opposite to the parton
reaction plane determined by the impact parameter $\vec{x}_T$, 
the same direction of the relative o.a.m.
This is essentially the manifest
of spin-orbital coupling in QCD. Ordinarily, the polarized
cross section along a fixed direction $\vec{n}_b$ vanishes when averaged
over all possible direction of the parton impact 
parameter $\vec{x}_T$. However, in non-central heavy-ion collisions
the local relative o.a.m. $L_y$ provides 
a preferred average reaction plane for parton collisions. This
will lead to a global quark polarization opposite to the
reaction plane of nucleus-nucleus collisions. This conclusion
should not depend on our perturbative treatment of parton scattering
as far as the effective interaction is mediated by the vector 
coupling in QCD.

Averaging over the relative angle between parton $\vec{x}_T$ 
and nuclear impact parameter $\vec{b}$
from $-\pi/2$ to $\pi/2$ and over $x_T$, one can obtain the
global quark polarization,
\begin{equation}
P_q = - \pi\mu\;p/4E(E+m_q)
\label{eq:pq}
\end{equation}
via a single scattering for given $E$. 
If we take $p=\Delta p_z/2=\vec{x}_T\cdot\hat{x}dp_z/dx$
before averaging, the result will be similar but
numerical evaluation is needed. In the limit $m_q=0$ and $p\ll \mu$,
one expects $P_q\sim -\Delta p_z/\mu$. Given an average 
range of interaction $\Delta x^{-1} \sim \mu\sim 0.5$ GeV
and $dp_0/dx=0.34$ GeV/fm for semi-peripheral ($b=R_A$) collisions 
(see Fig.~\ref{fig2}) at RHIC, $P_q \sim -0.3$.
Multiple scattering will further increase the polarization.

In nonrelativistic 
limit for massive quarks, $m_q \gg p,\mu$, 
\begin{equation}
P_q\approx - \pi \mu \; p/8m_q^2. \label{eq:nonrel}
\end{equation}
In the same limit, the spin-orbital coupling energy
is $E_{LS}=(\vec{L}\cdot\vec{S})(dV_0/dr)/(r\,m_q^2)$. Given the
range of interaction $r\sim 1/\mu$, $dV_0/dr \sim -\mu^2$
and $L\sim p/\mu$, $E_{LS}/\mu \sim - \mu p/m_q^2$ is just the
above quark polarization. 

The global quark polarization opposite to the reaction plane
will have many observable consequences 
in non-central heavy-ion collisions if an interacting QGP is formed.
One expects to see left-right asymmetry in hadron spectra at large rapidity 
similar to the single-spin asymmetry 
in $p+p$ collisions \cite{LM92,LB00,Troshin:1996hd,Collins:1992kk,Qiu:1998ia}.
Thermal photons, dileptons and final hadrons with spin will 
be similarly polarized. Since hadrons from the strong
decay of polarized resonances have angular distributions
that prefer the direction perpendicular to the resonances' 
polarization, they will result in an azimuthal asymmetry 
with respect to the reaction plane, similar to the asymmetry 
due to elliptic flow. In the following, we will 
discuss global hyperon polarization since it can be easily 
measured through the weak decay \cite{note}.

To demonstrate the robustness of the qualitative features 
of the predicted hyperon polarization due to global quark 
polarization, we consider several hadronization scenarios.
We consider first hadronization via parton recombination.
In this case, not only the spin of polarized quarks but also
the relative o.a.m. can contribute to the final hadrons'
polarization. Given hadron size $\Delta x$, the classical
estimate of the relative o.a.m. is $L_y \sim -(\Delta x)^2dp_z/dx$.
In the nonrelativistic quark model, however, constituent 
quarks in the ground state are all in the s-wave state. The
contribution of o.a.m. to hadron polarization resides in
the total angular momentum of the constituent quark.
%
Over all, the effective polarization of a constituent quark
should be proportional to that of the valence quark.
Alternatively, polarization of constituent quarks can
be similarly estimated, {\it e.g.} Eq.~(\ref{eq:nonrel}),
assuming massive constituent quarks as point-like particles.

We can categorize recombination into 
exclusive $qqq\rightarrow H$ and inclusive
$qqq\rightarrow H + X$ processes. The production
cross section of polarized hyperons from the exclusive 
recombination can be written as
\begin{equation}
\sigma_H^{\lambda}=
\sum_{\lambda_1,\lambda_2,\lambda_3}
|\langle q^{\lambda_1}_1 q^{\lambda_2}_2  q^{\lambda_3}_3
|H^{\lambda}\rangle |^2 \sigma_{q_1q_2q_3}
R_{q_1}^{\lambda_1}R_{q_2}^{\lambda_2}R_{q_3}^{\lambda_3},
\end{equation}
where $R_q^{\lambda}=(1+ \lambda P_q)/2$ is the quark
polarization probability. Extracting $R_H^{\lambda}$ from
$\sigma_H^{\lambda}\equiv \sigma_H R_H^\lambda$ that only
depends on quark polarization $P_q$, one can calculate the
recombination probability $R_H=R_H^{\uparrow}+R_H^{\downarrow}$
and the hyperon polarization $P_H=(R_H^{\uparrow}-R_H^{\downarrow})/R_H$.
Given baryons' SU(6) wavefunctions in the quark model \cite{LB00}, 
the polarization for strange ($P_s$) and non-strange quarks ($P_q$), 
one can obtain 
$P_{\Lambda}=P_s$, $R_\Lambda=3(1-P_q^2)$;
\begin{eqnarray}
P_{\Sigma}&=&(4P_q-P_s-3P_sP_q^2)/R_\Sigma, \
R_\Sigma=3-4P_qP_s+P_q^2; \nonumber \\
P_{\Xi}&=&(4P_s-P_q-3P_qP_s^2)/R_\Xi, \
R_\Xi=3-4P_qP_s+P_s^2; \nonumber \\
P_\Omega&=&2P_s(5+P_s^2)/R_\Omega, \ \ \ \ \ \ \ \ \ \ \  R_\Omega=6(1+P_s^2).
\nonumber
\end{eqnarray}
If $P_s\simeq P_q$ in the most likely case, we have
the same $P_H=P_q$ with $R_H=3(1-P_q^2)$ 
for $\Lambda$, $\Sigma$ and $\Xi$.

It is difficult to estimate hyperon polarization from
inclusive recombination. However, it could be
the dominant process for the bulk hadron production,
especially for large values of $P_q$. It is also required 
by entropy conservation. The polarization of produced hyperons
would be smaller than in the exclusive recombination but should
be proportional to $P_q$.

The extreme limit of inclusive recombination is fragmentation
of polarized quarks, $q \rightarrow H + X$.
Longitudinal polarization of hyperons in a similar process
$e^+e^-\to Z^0\to q\bar q\to \Lambda+X$ has been
measured \cite{lep} and can be explained \cite{liang} by assuming
that polarized hyperons contain the initial polarized 
leading quark in its SU(6) wavefunction. One can similarly
calculate hyperon polarization from fragmentation of
transversely polarized quarks and obtain,
\begin{eqnarray}
P_\Lambda&=& n_sP_s/(n_s+2f_s),
P_\Sigma=(4f_sP_q-n_sP_s)/3(2f_s+n_s),\nonumber \\
P_\Xi &=&(4n_sP_s-f_sP_q)/3(2n_s+f_s), P_\Omega =P_s/3,\nonumber
\end{eqnarray}
where $n_s$ and $f_s$ are the strange quark abundances relative to
up or down quarks in QGP and quark fragmentation, respectively.
Including production of unpolarized hyperons in the process, 
the effective hyperon polarization from fragmentation of polarized
quarks should increase with the fractional momentum $z_H=p_H/p_q$.

Because of the complication from hadronization, we cannot provide 
a model independent estimate of the final hyperon
polarization. If quark recombination is the dominant hadronization
mechanism, hyperons' polarization will be determined by the quarks'
polarization before hadronization. According to our estimate following
Eq.~(\ref{eq:pq}), this would be in the order of tens of a percent and
the hyperons' polarization would be around the same order.
We can also provide other
qualitative predictions of the global hyperon polarization $P_H$ 
in non-central heavy-ion collisions:
(1) Hyperons and their anti-particles are similarly polarized
along the same direction perpendicular to the reaction plane in 
non-central heavy-ion collisions.
(2) The global hyperon polarization $P_H$ vanishes in central collisions 
and increases almost linearly with $b$ in semi-central collisions.
(3) It should have a finite value at small $p_T$ and in the central 
rapidity region. It should increase with rapidity and eventually 
decreases and vanishes at large rapidities. 
(4) High $p_T \gg \Delta p_z\sim \mu L_0$ quarks should not be
polarized by parton scattering in the medium. However,
hyperon polarization should persist at moderate $p_T$ where
recombination of unpolarized high $p_T$ quarks with polarized
thermal quarks may still dominate. The polarization can be estimated
within a recombination model.
(5) Since hyperon's production planes are randomly oriented with
respect to the reaction plane of heavy-ion collisions, the
observed hyperon polarization in $p+A$ collisions should not
contribute to the global polarization as we have defined here,
except at large rapidity region where directed flow is
observed \cite{star-v1}. In this region, the non-vanishing
$\langle p_x\rangle$ can provide an average production 
plane $\vec{n}_H$ for hyperons.
According to the observed polarization pattern in $p+A$ \cite{PhData},
the global $P_{H}$ will be enhanced for $\Lambda$, $\Xi$, and $\bar\Xi$
and reduced for $\Sigma$ in large rapidity region. In addition,
one also expects $P_{\Lambda}>P_{\bar\Lambda}$ due to the directed flow.


In principle, hadronic interaction with a given direction of 
relative o.a.m. can also lead to global hyperon polarization.
Neglecting hyperon production in the hadronic
phase, the dominant hadronic processes involving hyperons will be
hyperon-$\pi$ scattering. The scattering amplitude in general
involves both scalar and vector channels \cite{Barros:2000iy}.
One can show that global hyperon polarization from the vector
channel is along while polarizations from the scalar channel and all 
the interferences are against the global quark polarization. Therefore,
polarizations from different channels in hyperon-$\pi$ scattering 
partially cancel each other. In addition, the transverse gradient 
of the longitudinal flow should be significantly reduced in the
hadronic phase due to prior strong transverse expansion 
in the reaction plane as demonstrated by the large elliptic
flow measured at RHIC. Therefore, the 
net effect of hadronic interaction should be small and should
not change the final hyperon polarization significantly, in
particular at large $p_T$. We will 
leave detailed study to future work.

In summary, produced partons are shown to have large local
relative o.a.m. in non-central heavy-ion 
collisions if quark-gluon plasma is formed. Parton
scattering with given relative o.a.m.
is shown to polarize quarks along the
same direction due to spin-orbital coupling. 
Such global quark polarization has many
measurable consequences in high-energy heavy-ion collisions.
Within different hadronization scenarios, we predict that hyperons
will be polarized along the opposite direction of the reaction 
plane. Effects of hadronic interaction are expected to be
small and would not change the qualitative feature of our prediction.

We thank S. Li and A. Majumder
for discussions. 
This work was supported by DOE
under No. DE-AC03-76SF00098,
NSFC No. 10175037 and No. 10440420018.


\section*{Erratum}

After the publication of the manuscript, we found
a factor of 2 missing from Eqs.~(8) and (9). Therefore the global quark
polarization in Eq.~(8) should be
\begin{equation}
P_q=-\pi\mu p/2E(E+m_q).
\end{equation}
In the non-relativistic limit for massive quarks, $m_q\ll p$,$\mu$,
the polarization becomes [Eq.~(9)],
\begin{equation}
P_q\approx -\pi\mu p/4m_q^2.
\end{equation}

We also made an error in the numerical estimate of the average longitudial momentum
difference within a typical interaction range $1/\mu$ in non-central $Au+Au$ collisions
at RHIC and consequently the quark polarization following Eq.~(8). The discussion
following Eq.~(8) should be replaced by:

In the limit $m_q=0$ and $p\gg \mu$, one expects $P_q=-\pi\mu/2\Delta p_z$.
Given $dp_0/dx=0.34$ GeV/fm for semi-peripheral ($b=R_A$) collisions at
RHIC and an average range of interaction $\Delta x^{-1}\sim \mu\sim 0.5$ GeV,
$\Delta p_z\sim 0.1$ GeV, which is much smaller than the average
transverse momentum transfer $\mu$. Therefore for a reliable estimate of 
the quark polarization at RHIC, one needs to go beyond the approximation
of small angle scattering.
  

\begin{thebibliography}{99}
\bibitem{note}
Experimental measurement of hyperon polarization relies on the
self-analyzing power of the parity violating weak decay, which
produces an angular distribution, $dN/d\cos\theta=1+\alpha P_H\cos\theta$,
for the decay products with respect to polarization direction, 
where $\alpha$ is a measured constant. It is clear that measurement 
of global hyperon polarization requires the determination of not only
the orientation but also the sign of the reaction plane. This can be
achieved through the measurement of the directed flow \cite{star-v1}
of produced hadrons at large rapidities. If multiple hyperons are
produced in the same event, one can also measure the global 
polarization via correlation between the spin of two hyperons.
\end{thebibliography}
\end{document}